\documentclass[a4paper,11pt]{article}
\pdfoutput=1
\usepackage[dvipsnames]{xcolor}
\usepackage[T1]{fontenc} 
\usepackage[toc,page]{appendix}
\usepackage{diagbox}
\usepackage{comment}
\usepackage{jheppub}
\usepackage{circledsteps}
\usepackage{subcaption}
\usepackage{natbib}
\setcitestyle{square,comma,numbers,sort&compress}
\usepackage{tabularx}
\usepackage{enumerate}
\usepackage{mathrsfs}
\usepackage{tabularx,booktabs}
\usepackage{caption}
\usepackage[compat=1.1.0]{tikz-feynman} 
\usepackage{cancel}
\usepackage{slashed}

\usepackage[section]{placeins}
\usepackage[capitalize]{cleveref}
\usepackage{enumitem}

\usepackage{simpler-wick}
\newcolumntype{Y}{>{\centering\arraybackslash}X}

\newcommand{\be}{\begin{equation}}
\newcommand{\ee}{\end{equation}}
\newcommand{\bea}{\begin{eqnarray}}
\newcommand{\eea}{\end{eqnarray}}

\definecolor{darkgreen}{rgb}{0,0.5,0}

\definecolor{myRED}{rgb}{0.8, 0.25, 0.33}

\definecolor{c1}{rgb}{0., 0.26, 0.5}
\definecolor{c2}{rgb}{0.9,0.648, 0.}
\definecolor{c3}{rgb}{0.75, 0., 0.}
\definecolor{c4}{rgb}{0., 0.6, 0.6}
\definecolor{c5}{rgb}{0.357, 0.35, 0.7}
\definecolor{c6}{rgb}{0.588, 0.6, 0.}

\newcounter{qnumber}

\begin{document}
\preprint{\hbox{UTWI-40-2024}}
\title{Indirect Detection of Hot Dark Matter}

\author[a]{Jeff A. Dror,}
\affiliation[a]{\sl Institute for Fundamental Theory, Physics Department,
University of Florida, Gainesville, FL 32611, USA}
\emailAdd{jeffdror@ufl.edu}
\author[b,c]{Pearl Sandick,}
\affiliation[b]{\sl Department of Physics and Astronomy, University of Utah, Salt Lake City, UT, 84112, USA}
\affiliation[c]{\sl Laboratoire Univers et Particules de Montpellier, CNRS \& Universit\'{e} de Montpellier, France}
\emailAdd{pearl.sandick@utah.edu}
\author[d]{Barmak Shams Es Haghi,}
\affiliation[d]{\sl Texas Center for Cosmology and Astroparticle Physics, Weinberg Institute for Theoretical Physics, Department of Physics, The University of Texas at Austin, Austin, TX 78712, USA}
\emailAdd{shams@austin.utexas.edu}
\author[a,b]{Fengwei Yang}
\emailAdd{fengwei.yang@ufl.edu}

\abstract{
Cosmologically stable, light particles that came into thermal contact with the Standard Model in the early universe may persist today as a form of hot dark matter. For relics with masses in the eV range, their role in structure formation depends critically on their mass. We trace the evolution of such hot relics and derive their density profiles around cold dark matter halos, introducing a framework for their {\em indirect detection}. Applying this framework to axions --  a natural candidate for a particle that can reach thermal equilibrium with the Standard Model in the early universe and capable of decaying into two photons -- we establish stringent limits on the axion-photon coupling $g_{a \gamma} $ using current observations of dwarf galaxies, the Milky Way halo, and galaxy clusters. Our results set new bounds on hot axions in the $\mathcal{O}(1-10)$\,eV range.
}

\maketitle
\flushbottom

\section{Introduction}
Particles with masses below the keV scale are ubiquitous in extensions of the Standard Model (SM) and can readily reach thermal equilibrium in the early universe. Direct search constraints for light particles demand that they interact feebly with the SM, implying they may be cosmologically stable. Despite their theoretical appeal, these light thermal relics cannot constitute cold dark matter (CDM) due to their suppressed small-scale density fluctuations, resulting from free streaming~\cite{Planck:2018vyg}. Instead, these particles can at most make up a small subcomponent of the matter-energy budget, known as {\bf hot dark matter} (HDM).

The properties of HDM particles ($ \chi $) are determined by their decoupling temperature, $T_d$. The distribution is a Bose-Einstein or Fermi-Dirac distribution until decoupling, at which time it ceases to be thermal as the universe expands. Since the particle momentum redshifts as $(1+z)^{-1}$ from the decoupling redshift $z_d$, the late-time phase space distribution is: 
\begin{equation} 
\label{eq:pdf_00}
f(p,z)=\frac{ 1  }{ \exp \left( \frac{ p }{ T _d } \frac{ 1+ z _d }{ 1 + z }     \right) \pm 1 }\,,
\end{equation} 
where the negative (positive) sign is for bosons (fermions). We define the {\em effective} temperature after decoupling to be $ T _\chi ( z ) \equiv T _d ( 1 + z ) / ( 1 + z _d ) $.

The cosmic energy density of HDM can be derived from the phase space distribution, yielding:
\begin{equation} 
 \Omega _{\chi} \simeq  0.008 ~g _\chi \left( \frac{m _\chi  }{ 5~ {\rm eV} } \right) \left( \frac{ T _{\chi, 0 } }{ 0.91~{\rm K} } \right) ^3 \left\{ \begin{array} {c}  1 \\ 3/4 \end{array} \right.\,,
\label{eq:Omega}
\end{equation} 
where the top (bottom) line corresponds to bosons (fermions), $g_\chi$ is the number of $\chi$ degrees of freedom, $m_\chi$ is the relic mass, and $T_{\chi,0}$ is the effective temperature today. For a hot relic that decoupled from the SM at temperatures well above 100~GeV in the standard cosmological history, $ T _{ \chi ,0 } = 0.91~ {\rm K} $. The value for $\Omega_{\chi} $ can be contrasted with the cosmic energy density of CDM, $\Omega_{\rm{CDM}} \simeq 0.27 $. 

Neutrinos are a well-studied example of HDM. Decoupling from the photon bath at a temperature of $\sim 1~{\rm MeV}$, their effective temperature today is approximately $ 1.9~{\rm K}$. Upon becoming non-relativistic, neutrinos infall into CDM halos. Similarly, other sub-keV thermalized particles, if present, would also contribute to the HDM component. Most efforts to detect HDM have focused on its gravitational influence, such as modifications to large-scale structure~\cite{Xu:2021rwg,Peters:2023asu,Dayal:2023nwi}. 

In this work, we propose an alternative approach: the {\bf indirect detection} of HDM through its potential decay channels. The approach parallels indirect direction strategies of CDM, targeting DM-rich regions with minimal astrophysical backgrounds to search for excess electromagnetic signals. Successful implementation requires calculating the density profiles of HDM, a subject that has received limited attention in the literature. 

We consider the observationally viable case where the energy density of HDM is much smaller than that of CDM. In this regime, two mechanisms govern the formation of HDM density profiles around CDM halos:
\begin{itemize}
    \item {\bf Concurrent collapse}: HDM collapses alongside CDM during halo formation and mergers, producing density profiles similar to those of CDM.
    \item {\bf Gravitational clustering}: HDM clusters within existing CDM halos via a late-time accretion process.
\end{itemize}
These mechanisms generate distinct density profiles: concurrent collapse produces cusp-like profiles, whereas gravitational clustering leads to flatter, more extended profiles. The dominant mechanism to drive the sensitivity of HDM indirect detection depends on the halo and particle masses. 

As a compelling application, we use this formalism to study axions ($ a$)~\cite{Peccei:1977hh,Peccei:1977ur,Weinberg:1977ma,Wilczek:1977pj} -- a light pseudoscalar particle with shift-symmetric interactions -- as HDM. Axions can decay into two photons ($ a \rightarrow \gamma \gamma $), producing a distinct electromagnetic signal. Our work extends previous HDM axion studies, which assumed efficient collapse akin to CDM~\cite{Ressell:1991zv, Grin:2006aw}. By incorporating the formation history of CDM halos, we robustly estimate the HDM density profile. Using existing infrared and optical telescope data, we constrain the axion-photon coupling in the $  1-10~{\rm eV}$ mass range. 

The paper is organized as follows. In \cref{sec:profiles}, we study the density profiles of $1-10~{\rm eV}$ mass HDM generated from concurrent collapse and gravitational clustering. Based on these density profiles, we develop the methodology for indirect detection of HDM in \cref{sec:indirect_detection}. Finally, we apply our work to study axions as HDM in \cref{sec:case_study}. We conclude in \cref{sec:conclusion}. This work uses natural units, in which $\hbar=c=k_B=1$. 

\section{Hot Dark Matter Density Profiles}
 \label{sec:profiles}
\subsection{Context and Key Concepts}
The density profiles of HDM are strongly influenced by the gravitational dynamics of CDM halos, which dominate the large-scale structure of the universe. During radiation domination, CDM density perturbations -- believed to be seeded by quantum fluctuations during inflation -- grow logarithmically inside the Hubble horizon. In the matter-dominated era, these perturbations grow linearly, eventually entering the non-linear regime, decoupling from the Hubble flow, and collapsing under their own gravity. This process leads to the formation of progenitor halos, which merge hierarchically to form larger structures. 

When the cosmic density of HDM is much smaller than that of CDM, HDM does not significantly alter the formation of CDM structures. However, HDM can still respond to the gravitational influence of CDM halos via {\em concurrent collapse} and {\em gravitational clustering}.

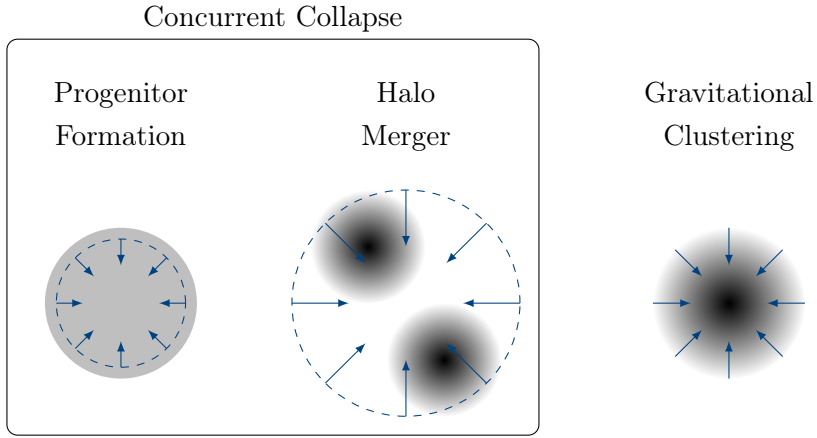
\begin{figure} 
\begin{center} \begin{tikzpicture} 
\draw[draw=none,inner color=gray,outer color=gray,fill opacity=0.5] (0,0) coordinate (A) circle (1cm);
\draw[draw=c1,dashed] (0,0) coordinate (A) circle (.85cm);
\foreach \th in {0,45,...,359} {\draw[c1,-latex] ({.85*cos(\th)},{.85*sin(\th)}) -- ({.5*cos(\th)},{.5*sin(\th)});}

\coordinate (merger) at (3.75,0);
\draw[draw=none,inner color=black,outer color=white] ($ (merger)+(.5,-.75)$) coordinate (g1)circle (.75cm);
\draw[draw=none,inner color=black,outer color=white] ($ (merger)+(-.5,.75)$) coordinate (g2) circle (.75cm);

\draw[draw=c1,dashed] (merger)  circle (1.5cm);
\foreach \th in {0,45,...,359} {\draw[c1,-latex] ($(merger)+({1.5*cos(\th)},{1.5*sin(\th)})$) -- ($(merger)+({.75*cos(\th)},{.75*sin(\th)})$);}

\draw[draw=none, inner color=black,outer color=white] (8,0) coordinate (g)circle (1cm);
\foreach[] \th in {0,45,...,359} {\draw[-latex,c1] ($ (g) + ({1*cos(\th)},{1*sin(\th)})$) -- ($(g)+ ({.5*cos(\th)},{.5*sin(\th)}) $);}
\draw[rounded corners] ($(A)+(-1.5,-1.75)$) rectangle ++ (7,5.25) coordinate[midway] (L);
\node at ($(L)+(0,2.9) $) {Concurrent Collapse};
\node at ($ (A) +(0,2.5) $) {\begin{tabular}{c}Progenitor \\  Formation \end{tabular}};
\node at ($ (g1)!0.5!(g2)  +(0,2.5) $) {\begin{tabular}{c}Halo \\ Merger \end{tabular}};
\node at ($ (g) +(0,2.5) $) {\begin{tabular}{c}Gravitational \\ Clustering  \end{tabular}};
  \end{tikzpicture}
\end{center}
\caption{Depiction of the two types of mechanisms leading to HDM density profiles around galaxies. The gray region depicts the CDM number density, while the blue arrows denote inflowing HDM particles. {\bf Left}: We depict {\em concurrent collapse} -- HDM undergoing gravitational instability during CDM halo formation with the blue dashed region denoting the unstable region. {\bf Right}: We depict {\em gravitational clustering} -- HDM infalling onto a static CDM halo. Depending on the halo mass and HDM phase space distribution, each process may be the most important driver of the dark matter density profile today.}
\label{fig:HDM}
\end{figure}

These processes depend on the HDM phase space distribution and the properties of CDM halos. In Fig.~\ref{fig:HDM}, concurrent collapse ({\bf Left}) and gravitational clustering ({\bf Right}) are illustrated, with the gray regions representing the CDM number density and the blue arrows indicating how HDM particles flow into the CDM halos.
 
For concurrent collapse, the region enclosed by the blue dashed line corresponds to a region that experiences gravitational instability.

To understand HDM halo formation qualitatively, we calculate the mean velocity of HDM after it becomes non-relativistic: 
\begin{align}
\label{eq:vfs}
\langle v\rangle(m _\chi , z ) &  = \frac{1}{ n _\chi ( z ) }\int \frac{d^3 {\mathbf{p}}}{(2\pi)^3} f(p, z) \frac{p}{m_\chi} \,,
\end{align} 
where $ n _\chi (z)$ is the number density of the relic at redshift $z$. For the homogenous phase space distribution of Eq.~\eqref{eq:pdf_00}, this evaluates to:
\begin{align} 
\label{eq:vavgtop}
\langle v\rangle(m _\chi , z ) & =(1+z)\frac{T_{\chi,0}}{ m_\chi} \frac{ \pi ^4 }{ 30 \zeta ( 3 ) }\left\{ \begin{array} {c}  1 \\ 7/6 \end{array} \right.\,, \\ 
& \simeq 10 ^{ - 3}  ~\frac{ 1 + z }{ 20 } \frac{ T _{ \chi ,0 } }{ 0.91 ~ {\rm K}}  \frac{ 10~{\rm eV} }{ m _\chi }\,.
\label{eq:vavg}
\end{align}
In Eq.~\ref{eq:vavgtop}, the upper line applies to bosons, and the lower line to fermions. 

This velocity can be compared to the {\bf critical velocity} $v_{\rm c}$, which is the {\em maximum} velocity required for HDM to gravitationally collapse into a halo. A rigorous treatment of gravitational instability requires studying the analytic structure of the perturbed phase space distribution. Carrying this out for the phase space distribution in Eq.~\eqref{eq:pdf_00} is beyond the scope of this work. Instead, we employ a qualitative estimate using the Jeans radius $R_{\rm J}$, which is derived under the assumption of a Maxwellian distribution \cite{2008gady.book.....B}: 
\begin{equation}
    R _{\rm J} \simeq \frac{1}{2} \sqrt{ \frac{ \pi v_{\rm c} ^2  }{ G \bar{\rho}_{\rm CDM} } }\,,
    \label{eq:RJ}
\end{equation} 
where $G$ is Newton's constant and $\bar{\rho}_{\rm CDM}$ is the background density of CDM. Note the appearance of $\bar{\rho}_{\rm CDM}$ instead of $\bar{\rho}_{\chi}$ in \cref{eq:RJ} is because the Jean's instability condition corresponds to the point when the perturbation decouples from the Hubble flow, and the Hubble parameter is determined by $\bar{\rho}_{\rm CDM}$ during matter domination.  Approximating the mass of the final CDM halo by $ \frac{4}{3}\pi\bar{\rho}_{\rm CDM} R^3_{\rm J} $, gives an expression for the critical velocity:
\begin{align} 
v_{\rm c}= \frac{6^{1/3}}{\pi^{5/6}}\left(  G ^3 \bar{\rho}_{\rm CDM} M ^2 _{\rm vir} \right)  ^{1/6}\,. \label{eq:v_circular}
\end{align} 
Evaluating this numerically, we find:
\begin{equation} 
v_{\rm c}(z) \simeq 3 \times 10 ^{ - 4} \left( \frac{ M_{\rm vir} }{ 10 ^{12} ~ M _{\odot} } \right) ^{1/3} \left( \frac{ 1 + z }{ 10} \right) ^{1/2} \,.
\end{equation}

\begin{figure}
\centering  
  \includegraphics[width=0.8\textwidth]{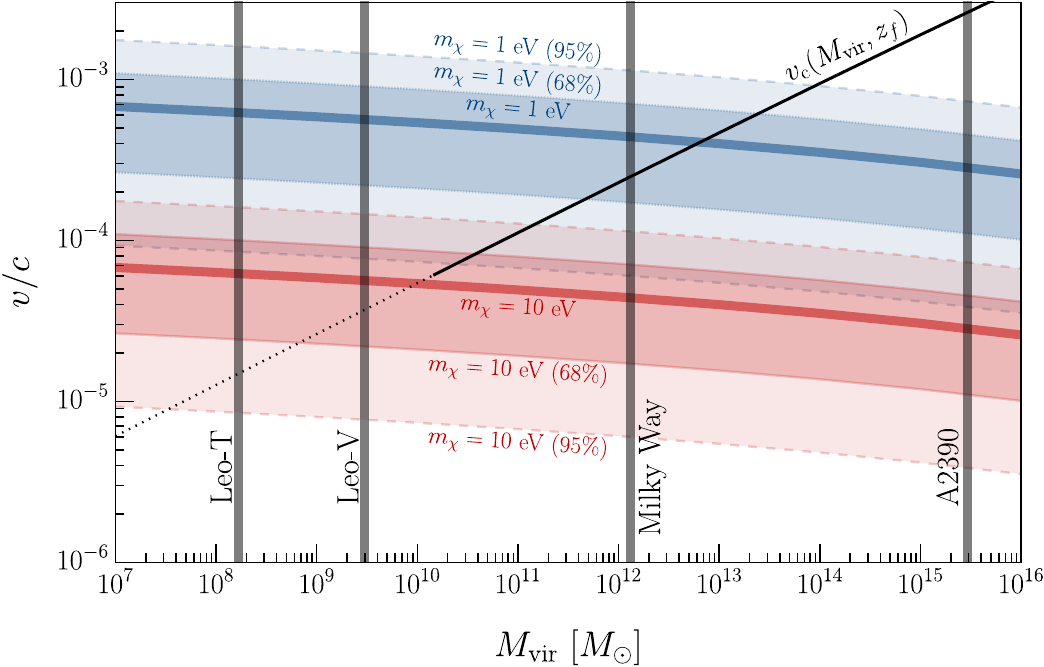}
 \caption{\label{fig:vesc} The comparison of the critical velocity of progenitor halos of various mass scales (black solid line) to the velocity of bosonic relic. The dark thick lines correspond to the mean velocity presented in Eq.~\eqref{eq:vavg} while the surrounding regions correspond to 68\% and 95\% of the phase space distribution at the halo formation time. We choose the current effective temperature of relic axions $T_{a,0}=0.91~{\rm K}$ and two axion masses: $1~{\rm eV}$ and $10~{\rm eV}$. The black dot-dashed line -- where $M_{\rm vir}\le 1.5 \times 10^{10}M_\odot$ -- requires extrapolating the fitting formula for the formation redshift derived in Ref.~\cite{10.1111/j.1365-2966.2011.19820.x}. The vertical gray lines label the halo mass of four CDM halos used to search for axion HDM in Section~\ref{sec:case_study}.} 
  \end{figure}

 \cref{fig:vesc} presents $v_{\rm c}$ as a function of halo mass at the formation redshift of the halo, $z_{\rm f}$, which we fix using the fitting function derived from simulations in Ref.~\cite{10.1111/j.1365-2966.2011.19820.x},~\footnote{The formation redshift -- also referred to as the assembly redshift -- is defined as the redshift at which the mass of the main progenitor of a halo, with a virial mass $M_{\rm vir}(z)$, first exceeds $M_{\rm vir}(z)/2$. This is determined directly from the halo's merger tree.
 
 Uncertainty in the best-fit halo formation redshift parameters from Ref.~\cite{10.1111/j.1365-2966.2011.19820.x} impacts the calculation of the critical velocity of CDM progenitor halos and the mean velocity of HDM at the redshift of halo formation, as $ v_{\rm c} \propto (1 + z)^{1/2}$ and $\langle v \rangle \propto (1 + z)$. We verified that varying these fit parameters within their 68\% confidence interval does not qualitatively affect the conclusions of this work. 
 
 Additionally, we note that the cosmological fits in Ref.~\cite{10.1111/j.1365-2966.2011.19820.x} were limited to $M_{\rm vir} \gtrsim 1.5 \times 10^{10} M_\odot$. In our analysis, however, we assume the fits apply to lower-mass halos as well.} 
 \begin{equation}
\label{eq:zform}
    \langle z_{\rm f }\rangle\simeq 1.1-0.22\log_{10}\left(\frac{M_{\rm vir}}{10^{12}~M_\odot}\right)\,.
\end{equation}
We compare the critical velocity to the velocity distribution of a bosonic HDM relic with a mass of 1~eV (blue) and 10~eV (red) at $z_{\rm f}$. The mean value -- as determined through Eq.~\eqref{eq:vavg} -- is given by the darker line, while the surrounding bands represent the range of velocities encompassing 68\% and 95\% of its total abundance. For comparison, we also show the halo masses of dwarf galaxies Leo-V and Leo-T, the Milky Way, and the galaxy cluster A2390 \cite{Hayashi:2022wnw,zoutendijk2021musefaintsurveyiiilarge,McMillan_2011,Newman_2013}, four systems we discuss in detail in the following sections. 

From Fig.~\ref{fig:vesc}, we observe that although the mean velocity of HDM may be large compared to $v_{\rm c}$, the spread in the phase space distribution of HDM suggests that a fraction of its population may be sufficiently slow-moving to gravitationally collapse. This slow-moving population undergoes concurrent collapse during the formation of CDM halos, a process which can take place during the formation of the first-generation progenitor halos or in halos formed from halo mergers. This population virializes akin to CDM, and, as such, we model the density profile as that of Navarro, Frenk, and White (NFW)~\cite{Navarro:1996gj}.

Additionally, the entire population can undergo gravitational clustering once a CDM halo becomes a static gravitational source. Following earlier work on gravitational clustering of the cosmic neutrino background~\cite{Singh:2002de, Ringwald:2004np}, here, we approximate this as a linear perturbation to the homogenous HDM phase space distribution surrounding CDM halos. This results in a distinct halo profile; in the inner radial region, the gravitational clustering profile plateaus to a fixed value, while the concurrent collapse profile is cusp-like. We find that in the outer radial region, the gravitational clustering profile extends further out than the concurrent collapse profile.

  The HDM halo density profile $\rho_\chi(r)$ as a function of the distance from the center of the halo, $r$, will be determined by both the concurrent collapse profile $\rho^{\rm cc}_\chi(r)$ and the gravitational clustering profile $\rho^{\rm gc}_\chi(r)$:
\begin{equation}
    \rho_\chi(r)=\rho^{\rm cc}_\chi(r)+\rho^{\rm gc}_\chi(r)\,.
\end{equation}
We derive expressions for both mechanisms, focusing on bosonic HDM. This choice is motivated by practical considerations; for fermionic relics, the Pauli-exclusion principle and Liouville's theorem restrict the density in halos from greatly exceeding the cosmic density~\cite{Tremaine:1979we}. 

\subsection{Concurrent Collapse} 
\label{sec:CC}
The fraction of HDM particles with velocity below the critical velocity of a CDM halo at formation behaves as CDM during structure formation. This {\em concurrent collapse} process results in a $ \chi $ density profile, $\rho^{\rm cc}_\chi(r)$, that follows the density profile of the CDM halo, $\rho_{\rm CDM} ( r )$, with an amplitude set by the fraction of $ \chi $ particles satisfying this criterion. The total mass of the to-be-collapsed HDM cloud, $M_\chi$, can be computed by estimating the number of the $\chi$ particles in the Jeans volume that may collapse into a halo of mass $M$:
\begin{equation}
\label{eq:M_chi}
    M_\chi=\frac{4\pi}{3}R_{\rm J}^3 m_\chi \bar{n} _\chi \varepsilon(M,z_{\rm f})\,,  
\end{equation}
where
\begin{equation}
\varepsilon (M, z_{\rm f}) \equiv \frac{g_\chi}{2\pi^2}\int_0^{m_\chi  v_{\rm c}(z_{\rm f})}dp p^2 f(p,z_{\rm f})\,,  
\end{equation}
denotes the fraction of $\chi$ particles with velocity below the critical velocity.~\footnote{When deriving the efficiency factor, we integrate over the unperturbed HDM distribution.} This estimate applies to the formation of a single CDM halo. The rich history of CDM halos requires evaluating this expression during every instance of gravitational instability during cosmic evolution. As progenitor halos form, they undergo mergers, leading to larger overdensities and creating opportunities for the gravitational collapse of surrounding HDM. Each CDM halo has a unique history -- characterized by their merger tree -- which is practically impossible to unravel. Nevertheless, it can be characterized statistically through semi-analytic methods. 

\subsubsection{Statistical Halo Formation}
We determine the history of a massive dark matter halo statistically by introducing the {\em progenitor halo mass function}, $n(M_1,z_1|M_2,z_2)$, which gives the number of progenitor halos of mass $M_1$ at redshift $z_1$ that undergo mergers and form halos of mass $M_2$ at redshift $z_2$. We calculate the total HDM mass that concurrently collapses with the CDM halo using
\begin{equation}
\label{eq:m_chi_tot}
    M_{\chi,{\rm tot}}=\int_{\infty}^{z_{\rm f}} dz_1\int_0^{M_{\rm vir}} dM_1 M_\chi(M_1)\frac{d}{dz_1}n\left(M_1, z_1|M_{\rm vir},z_{\rm f}\right)\,,
\end{equation}
where $M_\chi(M_1)$ is obtained from \cref{eq:M_chi} by determining the critical velocity $v_{\rm c}$ for a given progenitor halo of mass $M_1$; $dn/dz$ is the differential mass function, introduced to take into account collapses that occur within each redshift interval $[z_1,z_1+dz_1]$; and $z_{\rm f}$ is the halo formation redshift for a CDM halo with halo mass $M_{\rm vir}$ observed today.~\footnote{In our numerical results, we choose a maximum redshift of $10$ though this value has negligible influence on the final concurrent collapse population.} In deriving \cref{eq:m_chi_tot}, we use the progenitor halo mass function as a proxy for the merger mass function. This approximately corresponds to assuming every progenitor halo merges with another progenitor of a similar mass.

To derive $n(M_1,z_1| M_2,z_2) $ we employ the extended Press-Schechter formalism~\cite{Bond:1990iw}, as outlined in Ref.~\cite{2010gfe..book.....M}, 
\begin{equation}
    n(M_1,z_1|M_2,z_2)dM_1=\frac{M_2}{M_1^2}f_{\rm PS}(\nu_{12})\left|\frac{d\ln\nu_{12}}{d\ln M_1}\right|dM_1,
\end{equation}
where $\nu_{12} \equiv(\delta_1-\delta_2)/\sqrt{\sigma^2(M_1)-\sigma^2(M_2)}$, $\sigma^2(M)$ is the mass variance
\begin{equation}
    \sigma^2(M)\equiv\frac{1}{2\pi^2}\int_{0}^{1/R}P(k)k^2dk\,,
\end{equation}
$P(k)$ is the power spectrum of the density perturbations, and $M=6\pi^2\bar{\rho}_{\rm CDM}R^3$ is the mass enclosed in a volume $V=6\pi^2R^3$ with comoving mean matter density $\bar{\rho}_{\rm CDM}$ (see \cref{app:EPS} for further details), and 
\begin{equation}
    f_{\rm PS}(\nu)=\sqrt{\frac{2}{\pi}}~\nu e^{-\nu^2/2}\,, 
\end{equation}
is the Press-Schechter multiplicity function which represents the fraction of mass associated with halos within a unit interval of $\ln \nu$. 

The density of HDM particles that concurrently collapse with CDM at the halo formation time $z_{\rm f}$ is characterized as,
\begin{equation}
\label{eq:rho_cc}
   \rho^{\rm cc}_\chi(r) = \rho _{ {\rm CDM}} ( r )  \frac{M_{\chi,{\rm tot}}}{M_{\rm vir}}\,,
\end{equation}
where $M_{\chi,{\rm tot}}$ is obtained from \cref{eq:m_chi_tot}. Notably, the result $M_{\chi,{\rm tot}}$ obtained from \cref{eq:m_chi_tot} is dominated by the final mergers in the halo merger tree.

We briefly comment on the application of this formalism to the case of fermionic HDM, where the phase space is bounded due to a combination of Liouville's theorem and Fermi-Dirac statistics. The coarse-grained phase space distribution of fermionic HDM is,
\begin{equation}
\label{eq:f_chiCG}
    \frac{3\rho_{\chi}(r)}{4\pi v_{\rm esc}^3(r)m_\chi^4}\,,
\end{equation}
where $v_{\rm esc}(r)=\sqrt{2GM_{\rm vir}(<r)/r}$ is the escape velocity of the particle at the radial distance $r$ inside the dark matter halo. The maximum value of a Fermi-Dirac distribution is $g_\chi/ (2(2\pi)^3)$. This sets the upper bound of the density profile of {\it bound} fermionic HDM, 
\begin{equation}
    \rho_{\chi}(r)\le \frac{g_\chi}{3(2\pi)^2}m_\chi^4 v_{\rm esc}^3(r).
\end{equation} 
Due to the $1/r$ dependence in the inner region of an NFW halo, this bound is always violated at some distance for the density profiles of fermionic HDM as calculated in \cref{eq:rho_cc}.  

\subsection{Gravitational Clustering}
\label{sec:clustering}
If a HDM particle has a velocity exceeding the critical velocity of the CDM halo at its formation redshift, it will remain unbound to the halo. Nevertheless, as the universe expands, its velocity decreases and it will begin to cluster around existing CDM overdensities. The dynamics determining the resulting density profile are dramatically different than those of concurrent collapse. This {\em gravitational-clustering} process has been studied in the literature for HDM fermions in the context of the cosmic neutrino background~\cite{Brandenberger:1987kf,Singh:2002de,Ringwald:2004np,Ali-Haimoud:2012fzp,LoVerde:2013lta} and we extend the results for HDM bosons.

Gravitational potentials of dark matter halos formed in the late universe perturb the phase space distribution of HDM. We solve for the perturbation using the collisionless Boltzmann equation of the phase space distribution in the presence of the gravitational potential of a CDM halo (since the HDM density is small compared to CDM, we may neglect its contribution in the potential):
\begin{equation}
\label{boltzmanneq}
\frac{\partial{f ( t , {\bf r}, {\mathbf{p}} ) }}{\partial{t}}+\frac{d r _i }{dt}\frac{ \partial f  ( t, {\bf r} , {\mathbf{p}} ) }{ \partial r _i }+\frac{ d p _i }{ d t} \frac{\partial  f ( t, {\bf r} , {\mathbf{p}} ) }{ \partial  p _i }  =0\,,
\end{equation}
where we employ the Einstein summation notation, $t$ is cosmic time, and $\bf{r}$ ($ {\mathbf{p}} $) is the physical coordinate (momentum) vector of the HDM. The acceleration is given by Newton's law of gravitation, 
\begin{equation}
    \frac{d p _i }{dt}=-\frac{ \partial   \Phi _{\rm CDM}({\bf r})}{ \partial  r _i } \,,
\end{equation}
 We assume the CDM halo is decoupled from the Hubble flow and not merging, such that $\Phi_{\rm CDM}$ is time-independent. 

We solve the collisionless Boltzmann equation perturbatively, decomposing the phase space distribution of HDM into a homogeneous, isotropic component and a small correction, $f(q)+\delta f({\bf x}, {\bf q}, \tau)$, where ${\bf x}$, ${\bf q}$, and $\tau$ denote comoving position, comoving momentum, and conformal time respectively. These are related to physical position $\bf r$, physical momentum $\bf p$, and cosmic time $t$ through the scale factor $a(t)$:
\begin{equation}
    {\bf x}=\frac{{\bf r}}{a}\,,~{\bf q}\equiv am_\chi \frac{d{{\bf x}}}{d\tau}=a{\bf p}-m_\chi \dot{a}{\bf r}\,,~d\tau=dt/a\,,
\end{equation}
where $\dot{a}\equiv da/dt$. Substituting this decomposition into \cref{boltzmanneq} and transforming from physical to comoving quantities,~\footnote{To perform this transformation correctly, one needs to use the chain rule in partial time derivatives,
\begin{align} 
\left(\frac{\partial f}{\partial t}\right)_{\bf x,q}
=\left(\frac{\partial f}{\partial t}\right)_{\bf r,p}+\left(\frac{\partial f}{\partial {\bf r}}\right)\left(\frac{\partial {\bf r}}{\partial t}\right)_{\bf x}+\left(\frac{\partial f}{\partial {\bf p}}\right)\left(\frac{\partial {\bf p}}{\partial t}\right)_{\bf x,q}.\nonumber
\end{align}} we solve for the Fourier transform of the perturbation size,~\footnote{We define the Fourier transform of a function $ g ( {\mathbf{x}} ) $ as,
$$
    \tilde{g }({\bf k})=\int_{-\infty}^{\infty} d^3{\bf x}\,g ({\bf x})e^{-i{\bf k}\cdot {\bf x}}.
$$
Note that the vector Fourier variable $ {\mathbf{k}} $ is distinct from the momentum $ {\mathbf{q}} $ in the phase space distribution.}
\begin{equation}
\label{gilbert_eq}
\delta \tilde{f}({\bf{k}},{\bf{q}},\eta)=-\frac{im_\chi}{k^2}4\pi G\int_{\eta_0}^{\eta}d\eta'e^{-i{\bf{k}}\cdot{\bf{q}}(\eta-\eta')/m_{\chi}}a^4(\eta')\delta\tilde{\rho}_{\rm CDM}({\bf{k}},\eta'){\bf k } \cdot \nabla _{\bf q} f \,,
\end{equation}
where $d\eta=d\tau/a=dt/a^2$ is defined as a new time variable with $\eta_0$ its initial value, and $\delta\tilde{\rho}_{\rm CDM}$ is the Fourier transform of the CDM halo density profile. In obtaining \cref{gilbert_eq}, we have assumed the initial phase space distribution is described by the isotropic and homogenous phase space distribution given by Eq.~\eqref{eq:pdf_00} and took $\nabla _{\bf q} \delta f \ll \nabla _{\bf q}  f $.~\footnote{This latter approximation is valid on dimensional grounds, $\frac{\nabla_{\bf q} \delta f}{\nabla_{\bf q} f}\sim \frac{\delta f /\sigma_\chi}{f/\langle v\rangle}\sim\frac{\delta\rho_\chi/\sigma_\chi^4}{\bar{\rho}_\chi/\langle v\rangle^4}$, where $\sigma_\chi$ is the velocity dispersion in the halo, while $\langle v\rangle$ is the mean velocity obtained in \cref{eq:vavg}. Since only the relic particles with $\langle v\rangle\lesssim\sigma_\chi$ can fall into the gravitational well of dark matter halo, one expects $\frac{\nabla_{\bf q} \delta f}{\nabla_{\bf q} f}\ll \frac{\delta f}{f}$. See further discussion in Ref.~\cite{Singh:2002de}.} Note that even though the initial ``time'' $\eta_0$ is poorly defined, the density profiles we will derive are insensitive to its value.

We introduce the first-order perturbation to the comoving number density of HDM particles $\chi$ in Fourier space:
\begin{equation}
 \delta \tilde{ n} _\chi ( k , \eta ) = g_\chi\int_{-\infty}^{\infty}\frac{d^3 q}{(2\pi)^3} \delta \tilde{f}({\bf{k}},{\bf{q}},\eta)\,.
\label{eq:ntilde}
\end{equation}
Substituting the solution in \cref{gilbert_eq} into this expression and integrating by parts yields the first-order perturbation to the comoving number density of HDM:
\begin{equation}
\label{eq:master_eq}
\delta \tilde{n}_{\chi}(k,\eta)=\frac{2g_\chi Gm_{\chi}}{\pi k}\int_{\eta_0}^{\eta}d\eta' a^4(\eta')\delta\tilde{\rho}_{\rm CDM}(k,\eta')\int_0^{\infty}dq q\frac{\sin[kq(\eta-\eta')/m_{\chi}]}{e^{q/T_{\chi,0}}\pm1}\,.
\end{equation}
The inverse Fourier transform of this equation gives the perturbation of the $\chi$ number density as a function of radial distance from the center of the CDM halo.

 \cref{eq:master_eq} is our main result for gravitational clustering and we solve it numerically when presenting our results. It can be solved analytically when there is a significant hierarchy between $ k $ and the free-streaming wavenumber of $\chi$, 
\begin{equation}
     k_{\rm FS}\equiv\sqrt{\frac{4\pi G a^2(\eta)\bar{\rho}_{\rm CDM}}{\sigma_v^2}}, 
\end{equation}
where $\sigma_v=1.7~T_{\chi,0}/(a (\eta) m_\chi )$ is the $\chi$ velocity dispersion. We reserve analytic treatment of the equation to \cref{app:analy}. The result in each limit is given by the asymptotic expansion of \cref{eq:master_eq} in the limit of large and small momenta: 
\begin{equation}
\label{eq:delta_chi}
    \delta \tilde{n} _\chi(k,\eta)\simeq   \frac{{\bar n}_\chi}{\bar{\rho}_{\rm CDM}}\delta \tilde{\rho }_{\rm CDM}(k,\eta)
    \left\{\begin{array}{lr}1 & ~~k \ll k_{\rm FS} \\ \ln (2\Lambda)k_{\rm FS}^2/k^2 &~~ k \gg k_{\rm FS} \end{array}\right.\,,
\end{equation}
where $\Lambda\equiv kT_{\chi,0}(\eta-\eta_0)/m_\chi$. This result is applicable for bosons. For fermions, the result is identical with $\ln(2\Lambda) \rightarrow 1$.

The number density profile is obtained by taking the inverse Fourier transform of Eq.~\eqref{eq:delta_chi}. When $k_{\rm FS} r_{\rm vir} \ll 1 $ (which is typically the case), the integral is dominated by the $k \gg k_{\rm FS}$ region. The resultant $\chi$ number density profile is given as a convolution of a window function $W$ and the density contrast of CDM halo in coordinate space,
\begin{equation}
\label{eq:delta_chi_x}
    \delta n_\chi(x,\eta)=\frac{\bar{n}_\chi}{\bar \rho _{\rm CDM}} W(k_{\rm FS}x)*\delta \rho_{\rm CDM}(x,\eta),
\end{equation}
where $*$ denotes the convolution and $W(k_{\rm FS}x)\simeq \ln (2\Lambda)(k_{\rm FS}^2/x) $.
The presence of the window function smooths the $\chi$ density profile for scales smaller than $k_{\rm FS}^{-1}$.

Calculating the number density from \cref{eq:delta_chi_x}, and setting $x=0$, we find the number density of $\chi$ at the halo center:
\begin{equation}
\label{eq:n_gc_ana}
    n_\chi^{\rm gc}(x=0,\eta)=\bar{n}_{\chi,0}\times4\pi \ln (2\Lambda)k_{\rm FS}^2\frac{r_sr_{\rm vir}}{1+r_{\rm vir}/r_s}\frac{\rho_s}{\bar{\rho}_{{\rm CDM}}(\eta)}\,,
\end{equation}
where we have used the NFW profile for the CDM halo density profile,
\begin{equation}
    \rho_{\rm CDM}(r)=\frac{\rho_s}{(r/r_s)\left(1+r/r_s\right)^2}.
\end{equation}

If we neglect the merger tree, the gravitational clustering number density can be readily compared to that of concurrent collapse,
\begin{equation}
    n_\chi^{\rm cc}(x)=\frac{M_\chi}{M_{\rm vir}m_\chi}\rho_{\rm CDM}(x)\,.
\end{equation}
For a CDM halo with mass $M_{\rm vir}$ formed at $z_{\rm f}$, the HDM total mass derived from concurrent collapse is given by \cref{eq:M_chi}:
\begin{equation}
    M_\chi \simeq \frac{4\pi}{3}R_{\rm J}^3m_\chi \bar{n}_{\chi} \left[1+\frac{m_\chi^2v_{\rm c}^2}{\zeta(3)T^2_{\chi}(z_{\rm f})}\ln(1-e^{-m_\chi v_{\rm c}/T_{\chi}(z_{\rm f})})\right]\,,
\end{equation}
where $T_\chi(z)\equiv T_{\chi,0}(1+z)$. The quantity in square brackets is bounded from $0.51$ to unity and we neglect it in this rough estimate. 

In the $x\gtrsim r_s$ regime, $\rho_{\rm CDM}(x)\simeq\rho_s (r_s/x)^3$, and the ratio between the two profiles is: 
\begin{equation}
\label{eq:ratio_n}
\frac{n_\chi^{\rm cc}(x)}{n_\chi^{\rm gc}(x)}\simeq\frac{\mathcal{O}(1)}{4\pi}\frac{r_s}{k_{\rm FS}^2 x^3}, \quad (x \gtrsim r_s)\,.
\end{equation}
In deriving this result, we have used the fact that the gravitational clustering profile is flat with the extension to $\sim r_{\rm vir}$ and $r_{\rm vir}/r_s\gg 1$ (valid for all halo masses considered in this study). Eq.~\eqref{eq:ratio_n} exhibits a characteristic radius, $x \sim   (\lambda_{\rm FS}^2r_s)^{1/3}$. For distances smaller than this radius, concurrent collapse is the dominant mechanism driving the density profile, while for distances exceeding this radius, gravitational clustering becomes the dominant mechanism. We confirm this feature in Fig.~\ref{fig:naOfr}, introduced in the subsequent section. 

We conclude our discussion on gravitational clustering by remarking on three limitations of our treatment. Firstly, we assume the dark matter halo profile is spherically symmetric. Realistic halos may be significantly flattened. Additionally, we assume the halo is static since $z=3$, neglecting the influence of mergers during clustering. Since heavier halos induce more gravitational clustering and our results are insensitive to the initial time $\eta_0$ used in solving the collisionless Boltzmann equation, we do not expect a more realistic treatment to modify our estimate here significantly.~\footnote{We numerically check the uncertainty of the clustered number density by varying the initial redshift. We find the resultant density profile varies by a few percent when the initial redshift is changed from $3$ to $8$.} Finally, our analysis is strictly only applicable in the linear regime where $\delta n_\chi / {\bar n}_{\chi} \ll 1 $, but we extend it into the non-linear regime. While this cannot possibly capture all the dynamics present in gravitational clustering, this approach has been remarkably successful in studies of the cosmic neutrino background \cite{Singh:2002de,Ringwald:2004np} and we assume this continues to hold for bosonic relics. 

\subsection{Density Profiles}
\label{sec:densityprofiles}
\begin{figure}
\centering  
  \includegraphics[width=0.8\textwidth]{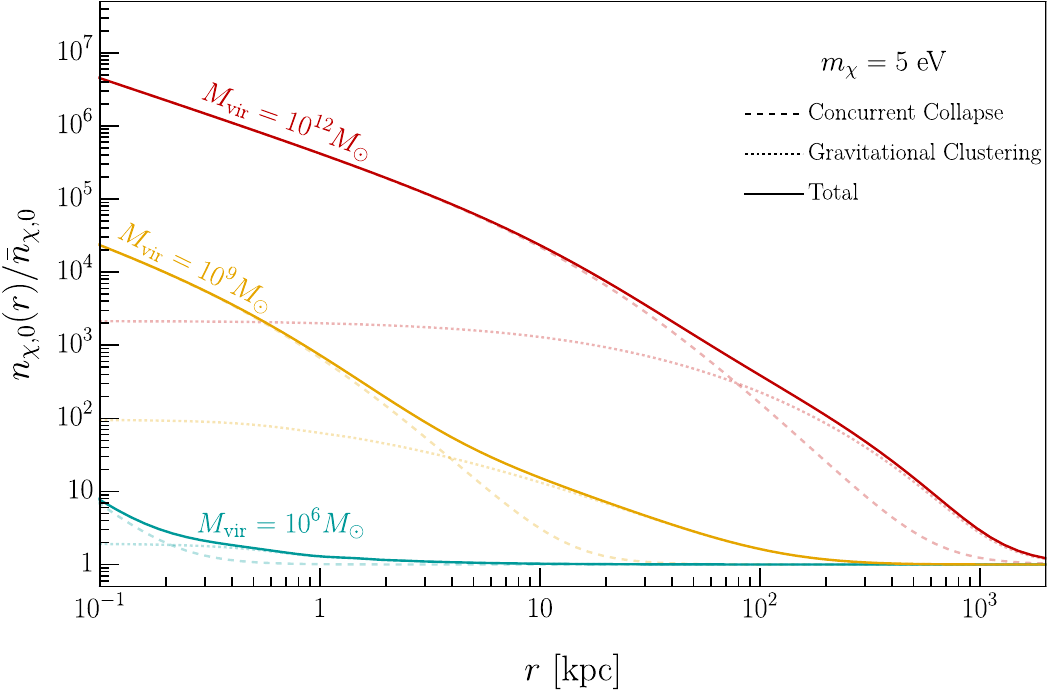}
 \caption{\label{fig:naOfr} The normalized number density profile of bosonic HDM (spin-$0$) in the CDM halo at redshift $z=0$ resulting from both concurrent collapse and gravitational clustering (solid lines). The concurrent collapse profile (dashed lines) and gravitational profile (dotted lines) are also shown. Note that the gravitational clustering profile extends further than the concurrent collapse profile. The halo mass $M_{\rm vir}$ is chosen to be $10^{6}~M_\odot$, $10^{9}~M_\odot$, and $10^{12}~M_\odot$, the mass of $\chi$ is $m_\chi=5$\,eV, and the current relic temperature is $T_{\chi,0}=0.91~{\rm K}$.}
\end{figure}

We calculate the density profiles of HDM using our key results, Eqs.~\eqref{eq:rho_cc} and ~\eqref{eq:master_eq}. \cref{fig:naOfr} shows the current number density profile -- normalized by the comoving number density of the unperturbed component -- for a HDM particle of $m_\chi=5~{\rm eV}$ in a CDM halo of $M_{\rm vir}=10^{6}~M_\odot$, $10^{9}~M_\odot$, and $10^{12}~M_\odot$. The solid lines represent the total number density profile of $\chi$, which is the sum of the concurrent collapse profile (dashed lines) and the gravitational clustering profile (dotted lines).

We emphasize two ubiquitous features of the density profiles. Firstly, the number density increases with the CDM halo mass, since a more massive halo has a higher escape velocity, which increases the available phase space for concurrently collapsing and gravitationally clustering particles. Secondly, the number density profile of HDM is dominated by concurrent collapse in the inner region, while gravitational clustering dominates in the outer region; the gravitational clustering profile extends further than the concurrent collapse profile. These features impact the expectations for indirect detection of HDM in nearby structures.

\section{Indirect detection of hot dark matter\label{sec:indirect_detection}}
Having calculated the density profiles of HDM, we use them to develop a strategy for its indirect detection. Decay of eV mass-scale HDM can produce excess photons in the infrared and optical bands, detectable by a range of telescopes. As with CDM indirect detection, a photon flux of frequency $\nu$ from the decay of HDM, $S(\nu)$, can be decomposed into a decay rate of HDM into photons, $\Gamma_{\chi\rightarrow\gamma}$, and an integral over its halo density profile, the ``$D$-factor,''
\begin{equation}
\label{eq:flux}
    S(\nu)\propto D \times\Gamma_{\chi\rightarrow\gamma}(m_\chi,g_{\chi\gamma})\,.
\end{equation}
The decay rate $\Gamma_{\chi\rightarrow\gamma} $ is fixed upon specifying the particle mass, $m_\chi$, and coupling, $g_{\chi\gamma}$. We will apply our formalism to the case of the axion in the following section. 

The $D$-factor, defined in detail in the next subsection, can be calculated using the results from \cref{sec:profiles}. We consider three different types of dark matter halos:
\begin{enumerate}[itemsep=0mm]
    \item dwarf galaxies Leo-V and Leo-T,
    \item the Milky Way, and 
    \item the galaxy cluster A2667 and A2390. 
\end{enumerate}
The approach to HDM indirect detection closely follows that of CDM with a novel set of $D$-factors. We calculate these for the different CDM seed halos, comparing the $D$-factors to those of CDM.

\subsection{The $D$-factor}
\begin{figure}
    \centering 
    \begin{tikzpicture}
    \coordinate (O) at (0,0);\node at ($(O)+(0,1) $) {Observer};
    \draw ($(O) $)-- ++ (0.5,.45) coordinate [near end] (A);
    \draw ($(O) $)-- ++ (0.5,-.45)coordinate [near end] (B);
    \draw[fill=gray, fill opacity=0.2] (A) to [out=-70,in=70] (B) -- (O) --cycle;
        \draw[draw=none,inner color=black!80!white,outer color=white] ($ (O)+(7,0)$) coordinate (g)circle (2.5cm);
        \draw[] ($(O)+(0,0) $) --($(g)+(0,0) $) node[midway,above] {$d$};
        \draw[-latex](g) --++ (-1,-1) coordinate (r) node[midway,below,xshift=0.cm,yshift=-0.1cm] {$r$} ;
        \draw[-latex] (O) -- (r) coordinate(X) node[below,midway] {$\ell$} ;
        \draw[dashed] ($(O)!0.4325!(g)$) to [out=-90,in=70] ($(O)!0.5!(r)$);
        \coordinate (Z) at ($(O)!0.4325!(g)+(0.2,-0.25)$); 
        \node at ($(Z)+(-0.0,-0.0) $) {$\theta$};
    \end{tikzpicture}
    \caption{Representation of the parameters defined in the differential $D$-factor. The observer is on the left while a observed dark matter halo is depicted on the right.}
    \label{fig:Dparameters}
\end{figure}
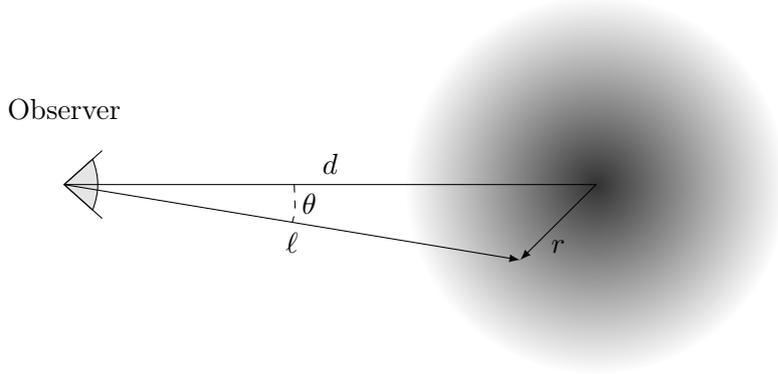
The differential $D$-factor is defined as an integral of the HDM density profile, 
\begin{equation}
\label{eq:diffD}
\frac{d D}{d\Omega}=\int_{\ell _{\rm min}}^{\ell _{\rm max}}d \ell \,\rho_{\chi}\left(r(\ell ,d,\theta)\right),
\end{equation}
where the coordinate $r$ is the distance from the center of the halo, $r=\sqrt{\ell ^2+d^2-2 \ell d\cos\theta}$, $d$ is the distance from the telescope to the center of the observed halo, $\ell$ is the distance along the line-of-sight, and $\theta$ is the angle between the line-of-sight and the direction of the center of the observed halo (see Fig.~\ref{fig:Dparameters}). The integral is performed along the line of sight, from $\ell _{\rm min}$ to $\ell _{\rm max}$, which are determined by the effective halo radius, $R$, and the geometry between the halo and the telescope:
\begin{equation}
    \ell _{\rm min}=d\cos\theta-\sqrt{R^2-d^2\sin^2\theta},~ ~\ell _{\rm max}=d\cos\theta+\sqrt{R^2-d^2\sin^2\theta}\,.
\end{equation}
The radius of a halo $R$ is poorly defined. We fix its value to the tidal radius $r_t$ for dwarf spheroidal galaxies inside the Milky Way halo,
\begin{equation}
   \frac{M_{\rm dSph}(r_t)}{r_t^3} =\frac{M_{\rm  MW}(d_{\rm dSph}-r_t)}{(d_{\rm dSph}-r_t)^3}\,,
\end{equation}
where $M_{\rm MW}(r)$ and $M_{\rm dSph}(r)$ are the halo masses enclosed within radius $r$ of the Milky Way halo and the dwarf spheroidal galaxy's halo, respectively, $d_{\rm dSph}$ is the distance from the center of the Milky Way halo to the center of the dwarf spheroidal galaxy's halo,
and the virial radius for the Milky Way and galaxy clusters. For Milky Way observations, where the telescope is inside the dark matter halo, the integration limits range from $\ell _{\rm min}=0$ to $\ell _{\rm max}=\infty$ (since the result is insensitive to distances far from the center of the Milky Way halo, we are free to neglect its virial radius).

\subsection{Search Procedure\label{sec:search_procedure}}
Any existing constraint on CDM indirect detection can be used to set a bound on HDM since the photon flux depends only on the product of the $D$-factor and the decay rate (see \cref{eq:flux}). By comparing the $D$-factor for the two cases (CDM and HDM): $D_{\rm CDM}$ and $D$, one can obtain a constraint on the $ \chi $-photon coupling for the HDM, $g_{\chi\gamma}$, by rescaling the constraint for the CDM, $g_{{\rm CDM}\gamma}$:\footnote{For the case when there are multiple halos in the observations, we generalize the right-hand-side of the equation to ratio of the sum of the $D$-factor of each halo, assuming the observation time of each halo is the same order of magnitude.}
\begin{equation}
\label{eq:rescaling}
\frac{\Gamma_{\chi\rightarrow\gamma}(m_\chi,g_{\chi\gamma})}{\Gamma_{{\rm CDM}\rightarrow\gamma}(m_{\rm CDM},g_{{\rm CDM}\gamma})}=\frac{D_{\rm CDM}}{D}\,.
\end{equation}
This procedure assumes that the outgoing photon spectrum of a HDM particle is identical to that of CDM. This is a good approximation when the experimental resolution is not sufficient to differentiate the HDM and CDM velocity distributions in the halo today. In the observations we use in our analysis, the resolution ranges between 1 part in a $10^2$ to 1 part in a $10^5$. These greatly exceed the width of CDM (1 part in $10^6$) and HDM in the ${\cal O}(1-10) ~{\rm eV} $ mass range we focus on in this work. 

We obtain the HDM $D$-factors using the methodology described in \Cref{sec:profiles}. We model the CDM halo density profile using the generalized NFW profile~\cite{1996MNRAS.278..488Z}, 
\begin{equation}
    \rho_{\rm CDM}(r)=\frac{\rho_s}{(r/r_s)^\gamma\left[1+(r/r_s)^\alpha\right]^{(\beta-\gamma)/\alpha}},
\end{equation}
where $\rho_s$ and $r_s$ are the halo characteristic density and radius, $\gamma$ and $\beta$ are the profile slopes in the inner and outer regions, respectively, and
$\alpha$ is the transition parameter to describe the slope transition from $\gamma$ for the inner region to $\beta$ for the outer region.

The NFW limit corresponds to $\{\alpha,\beta,\gamma\}=\{1,3,1\}$. In this case, the two model parameters $\rho_s$ and $r_s$ can be determined by the dimensionless concentration parameter $c\equiv r_{\rm vir}/r_s$ and the virial mass of the halo $M_{\rm vir}$:
\begin{equation}
\rho_s=\frac{\Delta_{\rm vir}\bar{\rho}_{m,0}a^{-3}}{3}\frac{c^3}{\ln(1+c)-c/(1+c)},
\end{equation}
\begin{equation}
r_s=\frac{r_{\rm vir}}{c}=\frac{1}{c}\left(\frac{3M_{\rm vir}}{4\pi\Delta_{\rm vir}\bar{\rho}_{m,0}}\right)^{1/3},
\end{equation}
where $\Delta_{\rm vir}$ denotes the average density of halo in terms of the average density of matter in the universe $\bar{\rho}_{m,0}$, i.e., $M_{\rm vir}\equiv (4\pi/3)\Delta_{\rm vir}\bar{\rho}_{m,0}r_{\rm vir}^3$. An analysis of approximately $5000$ halos with masses ranging from $10^{11}-10^{14}M_\odot$, finds an empirical formula for $c$:~\cite{Bullock:1999he}
\begin{equation}
\label{eq:cOfz}
c\simeq \frac{13}{1+z}\left(\frac{M_{\rm vir}}{10^{12}~M_{\odot}}\right)^{-0.13}\,.
\end{equation}

In this work, we set constraints on the HDM-photon coupling, $g_{\chi\gamma}$, using Eq.~\eqref{eq:rescaling} from observations by MUSE\,\cite{Regis:2020fhw}\footnote{For the MUSE observations one needs to further take into account the energy and angular response of telescopes, and obtains the $ D$ factor by doing the solid-angle integral of the differential $D$-factor weighted by a beam function $B(\Omega)$ is 
\begin{equation}
\label{eq:Dfac}
D= \int d\Omega\,\frac{d D}{d\Omega}B(\Omega)\,.
\end{equation}
We parameterize the beam function as a Gaussian with the full width at half maximum (FWHM) a function of frequency.}, WINERED\,\cite{Yin:2024lla}, JWST\,\cite{Janish:2023kvi}\footnote{We also note prior work on constraints on CDM axion decay from JWST in \cite{Roy:2023omw}.}, and VIMOS\,\cite{Grin:2006aw}\footnote{We note that Ref.~\cite{Grin:2006aw} studies both thermal axions and a more general scenario in which the eV-scale axions can make up the entirety of the dark matter.  We recast the constraints in Ref.~\cite{Grin:2006aw}, as reported in~\cite{AxionLimits}, on the latter scenario, relevant for CDM, to the HDM thermal axions studied here.}. Constraints from MUSE observations come from the dwarf galaxy Leo-T. Constraints from WINERED come from two dwarf galaxies, Leo-V and Tucana II. For simplicity, we consider only the most sensitive dwarf galaxy, Leo-V, and assume that the constraints derived in~\cite{Yin:2024lla} can be taken to be derived from Leo-V alone. As such, the constraints we report from Leo-V/WINERED should be considered an estimate.  We reserve the inclusion of the additional dwarf targets for future studies.  The constraints from JWST come from observations of the Milky Way dark sky.  Finally, VIMOS constraints come from two galaxy clusters, A2390 and A2667, and we include both as described below. 

Since the halo profiles measured from observations do not cover the full range of radii or redshifts, we use the empirical relation for $c$ given by \cref{eq:cOfz} to model the halo profiles in our analysis. 
Given this empirical relation between $c$ and $M_{\rm vir}$, the NFW profile depends solely on the parameter $M_{\rm vir}$ for a fixed $\Delta_{\rm vir}$. We choose $\Delta_{\rm vir}=200$ and model 
Leo-T's halo with $M_{\rm vir}=1.7\times10^{8}~M_\odot$ \cite{zoutendijk2021musefaintsurveyiiilarge}, 
the Milky Way's halo with $M_{\rm vir}=10^{12}~M_\odot$ \cite{McMillan_2011}, 
the A2667's halo with $M_{\rm vir}=2.0\times10^{15}~M_\odot$, and the A2390's halo with $M_{\rm vir}=2.9\times10^{15}~M_\odot$  \cite{Newman_2013}. 
The ultra-faint dwarf galaxy Leo-V halo is the faintest halo we study and may not be well described by an NFW halo. Following Ref.~\cite{Hayashi:2022wnw}, we model it with the generalized NFW profile parameters $\{\rho_s,\,r_s,\,\alpha,\,\beta,\,\gamma\}=\{5.0\times10^{-3}\,M_\odot{\rm pc}^{-3},\,6.3\,{\rm kpc},\,1.9,\,6.3,\,0.6\}$.
 
Before proceeding, we briefly discuss some subtleties related to our modeling of the CDM density profiles of the target objects, which introduces minor uncertainties in the constraints derived here. 
For Leo-T's halo, we compare the NFW profile adopted here with that determined by MUSE observations~\cite{zoutendijk2021musefaintsurveyiiilarge} ($\rho_s=2.1\times 10^8~M_\odot/{\rm kpc^3}$ and $r_s=0.28\,{\rm kpc}$), and find a relative difference of less than 10\% for radii greater than 1~kpc. For the Milky Way halo, the NFW profile adopted by Ref.~\cite{Janish:2023kvi} is that from Ref.~\cite{Cirelli:2010xx} ($\rho_s=0.18~\rm GeV/{\rm cm^3}$, and $\,r_s=24~{\rm kpc}$).  Our profile has a relative difference of less than 18\% across all radii. For A2390 and A2667, Ref.~\cite{Grin:2006aw} does not directly provide the best-fit values of $\rho_s$ and $r_s$ obtained in their analysis. We, therefore, assume an NFW profile as described above. We verified our profiles by comparing them to profiles derived from strong and weak gravitational lensing observations~\cite{Newman_2013}, 
and find a relative difference for all radii $r\le r_s$ of below 15\% and 18\% for A2390 and A2667, respectively. 
Finally, for Leo-V, the analysis in Ref.~\cite{Yin:2024lla} uses the profile from Ref.~\cite{Hayashi:2022wnw}, which we adopt here.  However, unlike Ref.~\cite{Hayashi:2022wnw}, 
we assume spherical symmetry of the halo, which introduces <1\% difference in the halo profile at any radius. 

\begin{figure}
\centering  
   \includegraphics[width=0.8\textwidth]{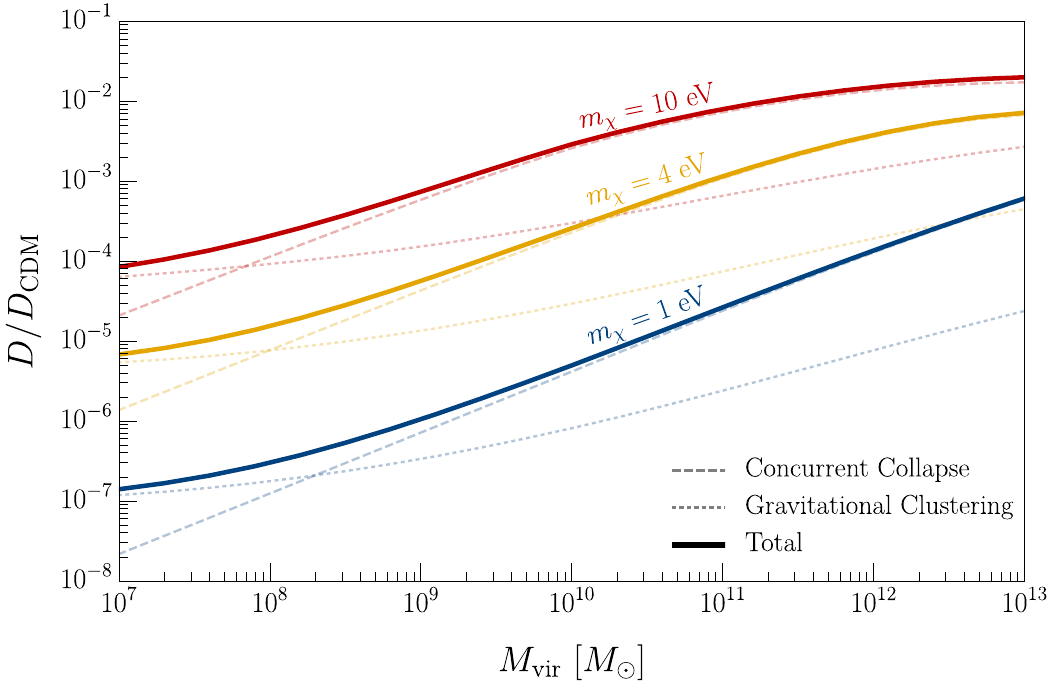}
 \caption{
 The $D$-factor ratio between spin-0 hot dark matter and the CDM halo profile as a function of halo mass, assuming an NFW profile with concentration parameter fixed by Eq.~\eqref{eq:cOfz}. We set the angle offset $\theta=0$, the field of view is $1'\times1'$, a uniform beam function, and the distance to the halo center is 4~Mpc. The $D$-factor ratios are shown in solid lines. The gravitational clustering contribution is shown in dashed lines, while the concurrent collapse contribution is shown in dot-dashed lines.} 
 \label{fig:Dfac_ratio} 
\end{figure}

The $D$-factors derived for HDM are compared with those of CDM in Fig.~\ref{fig:Dfac_ratio} as a function of halo mass. We present results for three bosonic HDM masses $m_\chi=1,\,4,\,10$\,eV, each corresponding to a fixed telescope angle offset $\theta=0$, a field of view $1'\times1'$, and the halo distance $d=4~{\rm Mpc}$. We use the empirical relation between the concentration parameter and halo mass in \cref{eq:cOfz} to model the NFW profile of CDM halos. We show the $D$-factor ratio using the total (solid lines) and separate contributions from concurrent collapse (dashed lines) and gravitational clustering (dotted lines). We observe that, while both concurrent collapse and gravitational clustering contributions increase with halo and HDM particle mass, the relative contribution of concurrent collapse is most important for heavier halos, while gravitational clustering is most important for lighter halos. Since the $D$-factor is a line-of-light integral of the density profile, it is sensitive to both the amplitude of the profile and the extension of the profile.

\section{Case Study: the Axion\label{sec:case_study}}
Axions provide a compelling motivation for conducting indirect detection searches for HDM. This section reviews the thermal history of axions and discusses their decay into two photons in the $1-10~{\rm eV}$ mass range. We also present novel constraints on the axion-photon coupling in the regime where axions decoupled from the SM bath well above the electroweak scale.

\subsection{Thermal history}
A generic axion arising from a global symmetry breaking at scale $f_a$ interacts with SM particles in the early universe with a rate of order $\sim T^3/f_a^2 $. This surpasses the Hubble parameter during radiation domination, $\sim T^2/M_{\rm pl}$, if the temperature in the early universe reached above $\sim f_a^2/M_{\rm pl}$. For $f_a \sim 10^{10} ~{\rm GeV}$, this corresponds to temperatures around 100 GeV. Such reheat temperatures are strongly motivated by the observed baryon-antibaryon asymmetry, making axions a prime HDM candidate. 

Their phase space distribution in the early universe is that of a redshifted Bose-Einstein distribution in Eq.~\eqref{eq:pdf_00}. Assuming no significant entropy dilution beyond that of the SM, the axion effective temperature today relates to the decoupling temperature as:
\begin{equation}
T_{a,0}=\left(\frac{g_{\star S} (T_0)}{g_{\star S}(T_d)}\right)^{1/3}T_{0}\,,
\label{eq:Ta0}
\end{equation}
where $g_{\star S}$ denotes the entropy degrees of freedom in the SM at a SM temperature $T$. For decoupling temperatures $T_d$ above a few hundred GeV, today's axion temperature is insensitive to $T_d$, yielding $ T _{ a ,0 } \simeq 0.91 ~{\rm K} $. We work in this limit and treat the decoupling temperature independently of axion-photon coupling strength.

\subsection{Decay Width}
The axion-photon interaction is described by: 
\begin{equation}
    \mathcal{L}_{\rm int}=-\frac{1}{4}g_{a\gamma}aF_{\mu\nu}\tilde{F}^{\mu\nu}\,,
\end{equation}
where $a$ is the axion field, $F_{\mu\nu}$ is the electromagnetic field strength tensor, and $\tilde{F}^{\mu\nu}=\frac{1}{2}\epsilon^{\mu\nu\alpha\beta}F_{\alpha\beta}$ its dual. This coupling induces axion decay into two photons, with a decay rate:
\begin{equation}
\Gamma_{a\rightarrow\gamma\gamma}=\frac{1}{64\pi}g_{a\gamma}^2m_a^3\,.
\end{equation} 
Thermal axions in the $1-10~{\rm eV}$ mass range become non-relativistic in the early universe, leading to the HDM density profiles discussed in \cref{sec:profiles}. Their decay channel offers a novel detection strategy, with emitted photons wavelengths given by:  
\begin{equation}
    \lambda_a = \frac{4\pi}{m_a} = 2479.68\,\text{nm}\left(\frac{\rm eV}{m_a}\right).
\end{equation}
Including redshift effects typical of galactic halos, photons from eV-scale axion decay fall within the infrared and optical wavelengths. 

For axions at rest, the decay produces a Dirac-delta energy spectrum peaked at frequency $\nu_a=\lambda_a^{-1}=m_a/(4\pi)$. More generally, the finite velocity dispersion introduces a small spectral width.  For HDM formed through concurrent collapse, virialized axions acquire the velocity dispersion of the corresponding CDM halos. For gravitational clustering, axions in the $1-10~{\rm eV}$ range have a mean velocity comparable to the 3D velocity dispersion of the CDM halo (see \cref{fig:vesc}).~\footnote{While the 3D velocity dispersion is not shown explicitly in the plot, it is of the same order as the critical velocity.} As such, we assume, in either case, that the width of the photon energy spectrum is within the frequency resolution of the observations. 

Using the $D$-factors calculated in Section~\ref{sec:profiles}, we constrain the axion-photon coupling using Eq.~\eqref{eq:rescaling}:
\begin{equation}
\label{eq:g_bound}
    g_{a\gamma}=\sqrt{\frac{D_{\rm CDM}}{D_a}}g_{{\rm CDM}\gamma}\,,
\end{equation}
where the $g_{{\rm CDM}\gamma}$ is the bound derived in the literature for CDM axions. 

\subsection{Results}
To constrain the axion-photon coupling under the HDM hypothesis, we applied the scaling relation from Eq.~\ref{eq:g_bound} using observational data collected with WINERED, MUSE, JWST, and VIMOS~\cite{2003SPIE.4837..910S,Zoutendijk_2020,Maiolino:2023wwm, Covone:2005uv,EJullo2008}. These instruments have previously been employed to set constraints under the CDM hypothesis~\cite{Regis:2020fhw,Janish:2023kvi,Roy:2023omw,Grin:2006aw,Yin:2024lla}. The raw data for the CDM axion constraints are obtained from \cite{AxionLimits}. For WINERED, 
the CDM search in Ref.~\cite{Yin:2024lla} incorporates data from two halos; they used observations of a second dwarf galaxy to reduce noise, thus the constraints come from a combination of observations.  Since it is not possible to separate the results from the two halos, for simplicity here we use exclusively the halo with the largest $D$-factor, Leo-V, and assume that the CDM $g_{a\gamma}$ constraints derived from each halo are the same.  As such, one should consider our results for WINERED to be an estimate.

\begin{figure}
\centering  
 \includegraphics[width=0.8\textwidth]{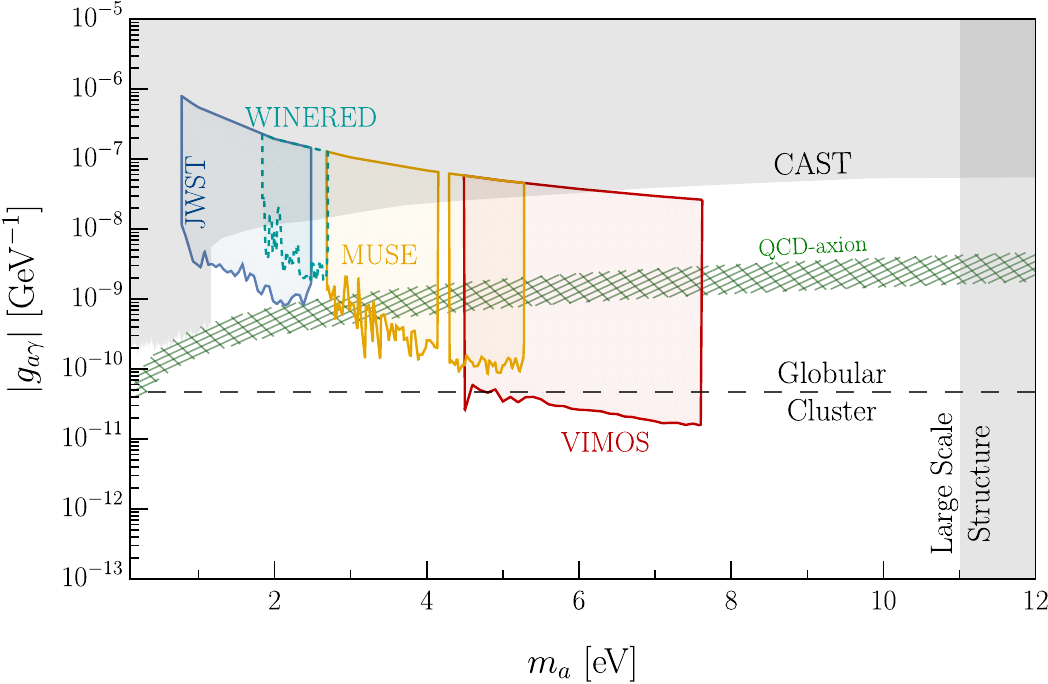}
 \caption{\label{fig:gaOfma} The indirect detection constraints (95$\%$ C.L.) on axion-photon coupling as a function of the axion mass for thermal axions, assuming a thermal relic decouples from the Standard Model bath above the weak scale. We show results by recasting data from the JWST telescope (blue)~\cite{Maiolino:2023wwm,Janish:2023kvi,Roy:2023omw}, the WINERED spectrograph mounted on the Magellan Clay telescope (teal, dashed)~\cite{2003SPIE.4837..910S,Yin:2024lla}, MUSE spectrograph installed on the Very Large Telescope (yellow)~\cite{Zoutendijk_2020,Regis:2020fhw}, and the VIMOS Ultra-Deep Survey (red)~\cite{Covone:2005uv,EJullo2008,Grin:2006aw}. Since the constraints assume the axion is cosmologically stable, the constraints require a maximum coupling for a fixed mass. The constraint from the CAST helioscope is shown in gray\,\cite{CAST:2007jps,CAST:2017uph}, while the constraints from modeling of star types within globular clusters -- more subject to astrophysical uncertainties -- is shown as the black dashed line~\cite{Dolan:2022kul}. We also show the constraint from searches for modifications to large-scale structure in gray~\cite{Xu:2021rwg}. The QCD-axion band is shown in green.}
\end{figure}

\cref{fig:gaOfma} presents the 95$\%$ C.L. HDM constraints derived under the assumption of a fixed decoupling temperature significantly exceeding 100 GeV from recasting data from the JWST telescope (blue)~\cite{Maiolino:2023wwm,Janish:2023kvi,Roy:2023omw}, the WINERED spectrograph mounted on the Magellan Clay telescope (teal, dashed)~\cite{2003SPIE.4837..910S,Yin:2024lla}, MUSE spectrograph installed on the Very Large Telescope (yellow)~\cite{Zoutendijk_2020,Regis:2020fhw}, and the VIMOS Ultra-Deep Survey (red)~\cite{Covone:2005uv,EJullo2008,Grin:2006aw}.
If the axion-photon coupling is sufficiently large that axions decay within the age of the universe, our formalism for deriving bounds becomes inconsistent. This imposes an upper limit on the coupling strength beyond which our constraints are no longer applicable. 
In addition, we overlay limits derived from the CAST helioscope experiment~\cite{CAST:2007jps,CAST:2017uph}, globular clusters star population ratios~\cite{Dolan:2022kul}, and large-scale structure gravitational HDM searches~\cite{Xu:2021rwg}. Notably, the HDM-based indirect detection constraints surpass those from CAST across most of the examined mass range and outperform globular cluster constraints for $m_a$ in the $4.5 - 7.6~{\rm eV} $ range.

For reference, we also show the QCD axion parameter space as a green band whose upper edge corresponds to the Kim-Shifman-Vainshtein-Zakharov (KSVZ) model~\cite{Kim:1979if,Shifman:1979if} and the lower edge corresponds to the Dine-Fischler-Srednicki-Zhitnitsky (DFSZ) model~\cite{Dine:1981rt,Zhitnitsky:1980tq}.  Note that the constraints derived here do not directly apply to the QCD axion parameter space, as QCD axions have additional couplings beyond the photon interactions that may keep it in thermal equilibrium below 100 GeV.

It is illuminating to compare the HDM axion constraints presented in this work with the CDM axion constraints. For CDM axions, JWST constrains $g_{\rm CDM \gamma} \lesssim 10^{-11}\,{\rm GeV}^{-1}$ \cite{Janish:2023kvi,Roy:2023omw}, WINERED provides $g_{\rm CDM \gamma} \lesssim 10^{-11}\,{\rm GeV}^{-1}$ \cite{Yin:2024lla}, VIMOS gives $g_{\rm CDM \gamma} \lesssim 2$\,-\,$5\times10^{-12}\,{\rm GeV}^{-1}$ \cite{Grin:2006aw}, and MUSE sets $g_{\rm CDM \gamma} \lesssim 10^{-13}\,{\rm GeV}^{-1}$ \cite{Regis:2020fhw}. The HDM axion constraints are relatively less stringent than the CDM axion constraints because of the relatively suppressed HDM $D$ factors (as shown directly in Fig.~\ref{fig:Dfac_ratio}). The upper bound on $g_{a\gamma}$ from VIMOS is the most stringent since it benefits from both a large halo mass and an observation frequency window sensitive to relatively larger axion masses.

Future telescope observations, particularly those targeting heavier halos with high HDM $D$-factors due to the larger critical velocities, hold the potential to significantly enhance sensitivity to axion HDM.

\section{Conclusion\label{sec:conclusion}}
In this work, we systematically study the indirect detection of HDM, a subcomponent of the total dark matter abundance expected in theories with high reheating temperatures. These temperatures allow light particles to thermalize in the early universe, setting the stage for HDM production. 

For eV-scale masses, HDM takes part in structure formation, resulting in HDM density profiles seeded by CDM halos. We identified two key mechanisms underlying HDM density profile formation: {\em concurrent collapse}, driven by gravitational instability in the HDM fluid during CDM halo formation, and {\em gravitational clustering}, an accretion process that occurs once the CDM halos become quasi-static. 

The decay of HDM produces photons observable with infrared and optical telescopes. By examining the relationship between CDM and HDM density profiles, we proposed a method to recast existing CDM indirect detection constraints as bounds on HDM. This approach exploits the fact that both constraints primarily depend on the line-of-sight integral over the halos profile and the decay rate. 

As a compelling application, we focused on eV-scale axions that decoupled from the thermal bath at temperatures well above 100 GeV. These axions decay into two photons, allowing us to derive novel constraints on the axion-photon coupling. 

Several simplifying assumptions in this work highlight opportunities for future study. For concurrent collapse, we assumed the gravitational instability of the HDM phase space distribution mirrors that of a Maxwell-Boltzmann distribution, with a critical velocity set by the Jeans scale. A more rigorous treatment would involve analyzing the perturbative properties of the phase space distribution, drawing from methods such as Landau's theory of instabilities~\cite{Landau:1946jc} as outlined in Binney and Tremaine's text on Galactic Dynamics~\cite{binney2011galactic} (Chapter 5). For gravitational clustering, we restricted our analysis to the linear regime; however, non-linear effects are likely significant and warrant further investigation.

Beyond axions, HDM indirect detection methods may also be applied to sterile neutrinos. These particles readily thermalize in the early universe and decay into photons through mixing with the SM neutrino ($N \rightarrow \nu \gamma$). However, detecting sterile neutrino decay is challenging, as the decay rate scales as the fifth power of the sterile neutrino mass. Exploring alternative decay channels or enhancing observational techniques may improve prospects for testing the sterile neutrino hypothesis.

As telescope observation time increases and spectroscopic sensitivity improves, large regions of the parameter space for eV-scale HDM will become accessible. These technological improvements promise to deepend our understanding of the light particle spectrum and their broader role in shaping our universe. 

\section*{Acknowledgments}
We thank Marco Cirelli, Paolo Gondolo, Jason Kumar, Julien Lavalle, Chung-Pei Ma, Caio Nascimento, Paul R. Shapiro, Elisa M. Todarello, Wen Yin, Yue Zhao, and Zheng Zheng for  valuable discussions. B.S.E and F.Y. would like to thank the Department of Physics and Astronomy, University of Utah, where part of the work was completed. P.S.~is supported by NSF grant PHY-2412834. F.Y. is supported by the U.S. Department of Energy under Award No. DE-SC0022148. B.S.E. acknowledges support by the U.S.\ Department of Energy, Office of Science, Award Number DE-SC-0022021. The research of JD is supported in part by the U.S. Department of Energy grant number DE-SC0025569.

\appendix
\section{Extended Press-Schechter formula\label{app:EPS}}
The halo mass function is obtained by studying the statistical properties of the collapsed regions which originates from the linear overdensity field $\delta({\bf x})\equiv\delta\rho_{\rm CDM}({\bf x})/\bar{\rho}_{\rm CDM}$ at some early time, where $\delta\rho_{\rm CDM}({\bf x})$ is the density perturbation of cold dark matter at position ${\bf x}$ and $\bar{\rho}_{\rm CDM}$ is the average cold dark matter density in the universe. Given the Gaussian populated random field $\delta({\bf x})$, one can statistically count the fraction of regions that are to be collapsed at some late time. Also, depending on how large the radius of the collapsed sphere is, one can further determine the mass of the collapsed object. By postulating the probability of the collapse of a given halo mass is equal to this fraction, one can define the halo mass function. Press and Schechter \cite{Press:1973iz} provide a formalism to analytically calculate the halo mass function, but in their formalism, there is a fudge factor that needs to be put by hand. Alternatively, Ref.~\cite{Bond:1990iw} used the excursion set formalism to derive the same halo mass function without the fudge factor, so-called extended Press-Schechter (EPS) formalism. The excursion set formalism treats the smoothed random field $\delta_s({\bf x})$ and its variance as a point in the trajectory that has an excursion and studies when the trajectory can pass through the critical threshold such that the collapse can happen. In this appendix, we summarize how to derive the progenitor halo mass function from the EPS formalism \cite{Bond:1990iw}.  

Consider the random Gaussian overdensity field $\delta({\bf x},t)$. Denote the extrapolation of $\delta({\bf x},t)$ to the present time, $\delta({\bf x})=\delta({\bf x},t)/D(t)$, where $D(t)$ is the growth factor, given by the scale factor $a(t)$ in the matter-dominated era. In the spherical collapse model, collapse happens when $\delta({\bf x},t)>\delta_c \simeq 1.69$, or equivalently, $\delta({\bf x})>\delta_c/D(t)$. To properly estimate the mass enclosed in the collapsed spherical region, Press \& Schechter \cite{Press:1973iz} chose a window function $W({\bf x})$ to define a smoothed overdensity within a characteristic radius $R$. In the excursion set formalism, a sharp $k$-space filter sets the window function, $\tilde{W}(kR)=\Theta(1-kR)$, and the smoothed overdensity field is then
\begin{equation}
\label{eq:delta_s}
    \delta_s({\bf x};R)=\int _{-\infty} ^{\infty} d^3{\bf k}\tilde{W}(kR)\delta_{\bf k}e^{i{\bf k}\cdot{\bf x}}\,,
\end{equation}
where $1/R$ is the size of the top-hat filter in $k$-space, and $\delta_{\bf k}$ are the Fourier modes of $\delta({\bf x})$. The mass variance of the smoothed field is
\begin{equation}
    \sigma^2(M)=\langle\delta_s^2({\bf x})\rangle=\frac{1}{2\pi^2}\int_{0}^{1/R}P(k)k^2dk,
\end{equation}
where $P(k)$ is the power spectrum of the density perturbations. For a $k$-space top-hat filter, the normalization of $W({\bf x})$ is conventionally chosen such that $M=6\pi^2\bar{\rho}_{\rm CDM}R^3$ is the mass enclosed in a volume $6\pi^2R^3$ with comoving mean matter density $\bar{\rho}_{\rm CDM}$~\cite{2010gfe..book.....M}. In this study, we use the linear power spectrum because we are only interested in the progenitor halo scale, where $k_c$ is not in the non-linear regime of the matter power spectrum. We model the linear power spectrum by 
\begin{equation}
    P_{\rm lin}(k,t)=P_0(k)D^2(t)T^2(k),
\end{equation}
where the adiabatic initial power spectrum $P_0(k)$ is related to the dimensionless power spectrum, $\Delta^2(k)=(k^3/2\pi^2)P_0(k)$, and
\begin{equation}
    \Delta^2(k)=\delta_H^2\left(\frac{k}{H_0}\right)^{n_s+3},
\end{equation}
with $\delta_H=3.5\times10^{-5}$,\footnote{We determine the spectrum amplitude by requiring the variance at $R=8~{\rm Mpc}/h$, where $h \simeq 0.67$, is given by $\sigma_8=0.8$.} $n_s=0.965$, $H_0$ the Hubble parameter today,
the transfer function $T$ can be approximated with the BBKS form \cite{Bardeen:1985tr} for an adiabatic initial power spectrum,
\begin{equation}
    T(k)\simeq \frac{\ln(1+1.94 q)}{1.94q}[1+3.22q+(13.3q)^2+(4.52q)^3+(5.55q)^4]^{-1/4},
\end{equation}
where $q\equiv  k/(10~{\rm Mpc})^{-1}$. 

Using the smooth field in \cref{eq:delta_s}, one can show that the change of $\Delta \delta_s$ corresponding to an increase from $k_{\rm c}$ to $k_{\rm c}+\Delta k_{\rm c}$ is a Gaussian variable with variance $\langle\Delta \delta_s\rangle=\sigma^2(k_{\rm c}+\Delta k_{\rm c})-\sigma^2(k_{\rm c})$, where $k_{\rm c}\equiv1/R$. So the distribution of $\Delta \delta_s$ is independent of the value of $ \delta_s({\bf x};k_{\rm c})$. When 
$k_{\rm c}$ is increased, one can describe the value of $ \delta_s({\bf x};k_{\rm c})$ by a Markovian random walk. By studying the probability when the trajectory of $(\sigma^2,\delta_s)$ with the origin at $(\sigma_2^2,\delta_2)$ executes the first up-crossing of the threshold $\delta_s=\delta_1$, with $\delta_{1,2}\equiv\delta_{\rm c}/D(t_{1,2})$ and $\sigma_{1,2}\equiv\sigma^2(M_{1,2})$, one obtains the first up-crossing function:
\begin{equation}
f_{\rm FU}(\sigma_1^2,\delta_1 | \sigma_2^2,\delta_2)=\frac{1}{\sqrt{2\pi}} \frac{\delta_1-\delta_2}{(\sigma_1^2-\sigma_2^2)^{3/2}}  \exp\left[-\frac{(\delta_1-\delta_2)^2}{2(\sigma_1^2-\sigma_2^2)}\right],
\end{equation}
where the first up-crossing happens for the first time when the trajectory of $(\sigma^2,\delta_s)$ goes from below the threshold $\delta_s<\delta_1$ to above the threshold $\delta_s>\delta_1$ while increasing the variance $\sigma^2$. Given the halo mass $M_2$ observed at time $t_2$, the fraction of the progenitor halos with mass $M_1$ formed at time $t_1$ is denoted as the {\it progenitor halo mass function}, which is determined by the first up-crossing function:
\begin{equation}
    n(M_1,t_1|M_2,t_2)dM_1=\frac{M_2}{M_1}f_{\rm FU}(\sigma_1^2,\delta_1 | \sigma_2^2,\delta_2)\left|\frac{d\sigma_1^2}{dM_1}\right|dM_1=\frac{M_2}{M_1^2}f_{\rm PS}(\nu_{12})\left|\frac{d\ln\nu_{12}}{d\ln M_1}\right|dM_1,
\end{equation}
where $\nu_{12}=(\delta_1-\delta_2)/\sqrt{\sigma_1^2-\sigma_2^2}$ and $f_{\rm PS}(\nu)=\sqrt{\frac{2}{\pi}}\nu \exp(-\nu^2/2)$ is the Press-Schechter multiplicity function.

\section{Analytical Understanding of Gravitational Clustering\label{app:analy}}
In the main text, we used an analytic formula for the perturbed number density due to gravitational clustering (Eq.~\eqref{eq:delta_chi}). In this appendix, we derive this result starting from~\cref{eq:master_eq}. 

We simplify the discussion, we introduce the dimensionless quantities, $x\equiv q/T_{\chi,0}$, $\xi\equiv(\eta-\eta')/\eta$. The master equation becomes, 
\begin{eqnarray}
   \delta\tilde{n}_\chi(k,\eta)
&=&\frac{2g_\chi Gm_{\chi}T_{\chi,0}^2}{\pi k}\int_{0}^{1-\eta_0/\eta}\eta d\xi a^4(\eta(1-\xi))\delta\tilde{\rho}_{\rm CDM}(k,\eta(1-\xi))\nonumber\\
&\times&\int_0^{\infty}dxx\frac{\sin(kxT_{\chi,0}\eta\xi/m_{\chi})}{e^x\pm1}\,.
\end{eqnarray}
One can approximate the above double integral by considering two regions when $kT_{\chi,0}\eta\xi/m_{\chi}$ $\ll 1$ (small $k$ region) and $kT_{\chi,0}\eta\xi/m_{\chi}\gg 1$ (large $k$ region). To better understand the meaning of the scale, we define the free-streaming wavenumber of hot relics $\chi$,
\begin{equation}
k_{\rm FS}\equiv\sqrt{\frac{4\pi Ga^2(\eta)\bar{\rho}_{\rm CDM}}{\sigma_v^2}},
\end{equation}
where the thermal velocity dispersion of $\chi$ is
\begin{equation}
   \sigma_v^2\equiv \langle v_{\chi}^2\rangle-\langle v_{\chi}\rangle^2,
\end{equation}
where the averages are derived from the unperturbed phase space distribution function. One finds $\sigma_v\approx1.7\frac{T_{\chi,0}}{m_\chi a(\eta)}$ for both fermionic and bosonic cases.~\footnote{In Ref.~\cite{Nascimento:2023ezc}, they define $\sigma_v^2=\int dv v^2 f(v,\eta)/\int dv f(v,\eta)$, which gives $\sigma_v=\sqrt{\frac{3\zeta(3)}{\ln(4)}}\frac{T_{\chi,0}}{m_\chi a(\eta)}\approx 1.6\frac{T_{\chi,0}}{m_\chi a(\eta)}$ for the fermionic case.} 
By substituting the thermal velocity dispersion and using $H^2=\frac{8\pi G}{3}\bar{\rho}_{\rm CDM}$, $H=\frac{2}{3}t^{-1}$, and $\xi\eta=\eta-\eta'=\int dt/a^2(t)\propto t^{-1/3}$ if $\eta\gg\eta'$ in the matter dominated era, we find $k_{\rm FS}\sim m_\chi/(T_{\chi,0}\eta\xi)$. So, equivalently, we can consider the region when $k\ll k_{\rm FS}$ as a small $k$ region and the region when $k\gg k_{\rm FS}$ as a large $k$ region.

In the small $k$ region, we can approximate the $x$-integral by using $\sin(kxT_{\chi,0}\eta\xi/m_{\chi})\approx kxT_{\chi,0}\eta\xi/m_{\chi}$:
\bea
\delta\tilde{n}_\chi(k,\eta)&\simeq&\frac{2g_\chi GT_{\chi,0}^3\eta^2}{\pi }\int_{0}^{1-\eta_0/\eta}\xi d\xi a^4(\eta(1-\xi))\delta\tilde{\rho}_{\rm CDM}(k,\eta(1-\xi)) \int_0^{\infty}\frac{x^2dx}{e^x\pm1}. \nonumber\\
\eea
The $x$-integral gives $2\zeta(3)$ in the fermionic case, while it gives $3\zeta(3)/2$ in the bosonic case. The important consequence is that the small-$k$ result is independent of the mass of relic particles $\chi$. By further doing the $\eta$-integral, one can find the density contrast of clustered relic particles follows the density contrast of CDM in the halo, $\tilde{\delta}_\chi(k,\eta)\equiv\frac{\delta\tilde{n}_\chi(k,\eta)}{\bar{n}_{\chi,0}}\approx\tilde{\delta}_{\rm CDM}(k,\eta)$, where $\tilde{\delta}_{\rm CDM}(k,\eta)\equiv\frac{\delta\tilde{\rho}_{\rm CDM}(k,\eta)}{\bar{\rho}_{\rm CDM}}$ \cite{Nascimento:2023ezc}.

In the large $k$ region, the $x$-integrand is highly oscillating which makes the integral zero, except that when $\xi\simeq0$. In this case, one can approximate the double integral by doing the Taylor expansion around $\xi=0$. The leading order term is, therefore,
\begin{align}
\label{lin_1}
\delta\tilde{n}_\chi(k,\eta)&\simeq\frac{2g_\chi Gm_{\chi}T_{\chi,0}^2\eta}{\pi k} a^4(\eta)\delta\tilde{\rho}_{\rm CDM}(k,\eta)\int_0^{\infty}\frac{xdx}{e^x\pm1}\int_{0}^{1-\eta_0/\eta} d\xi\sin[kxT_{\chi,0}\eta\xi/m_{\chi}] \nonumber\\
&=\frac{2g_\chi Gm_{\chi}^2T_{\chi,0}}{\pi k^2}  a^4(\eta)\delta\tilde{\rho}_{\rm CDM}(k,\eta)I_\chi,
\end{align}
where we define the phase space integral by,
\begin{equation}
I_\chi\equiv\int_0^{\infty}\frac{dx}{e^x\pm1}[1-\cos(kxT_{\chi,0}(\eta-\eta_0)/m_{\chi})]\,.
\end{equation}
For the fermionic case, this integral is well approximated by neglecting the second term, which gives $I_\chi=\ln(2)$. However, in the bosonic case, the integral has a divergent point at $x=0$, and the second term is necessary when carrying out the integral. The result is given by the Harmonic number $I_\chi={\rm Re}(H_{2\Lambda i})=\frac{1}{2}(H_{2\Lambda i}+H_{-2\Lambda i})$, where $\Lambda=kT_{\chi,0}(\eta-\eta_0)/m_{\chi}$ and $i$ is the imaginary unit. Since $kT_{\chi,0}\eta\xi/m_{\chi}\gg 1$, $\Lambda\gg 1$, and therefore, the integral can be approximated by the asymptotic value of ${\rm Re}(H_{2\Lambda i})\sim \ln(2\Lambda)+\gamma\approx \ln(2\Lambda)$, where $\gamma\approx0.577$ is the Euler–Mascheroni constant. 

By using $k_{\rm FS}$, one can further simplify eq.\,(\ref{lin_1}) in the bosonic case to
\begin{equation}
    \tilde{\delta}_{\chi}(k,\eta)\simeq\ln (2\Lambda) \frac{k^2_{\rm FS}}{k^2}\tilde{\delta}_{\rm CDM}(k,\eta)\,,
\end{equation}
where $\Lambda\equiv kT_{\chi,0}(\eta-\eta_0)/m_\chi$. The expression for fermions is identical with setting $\ln (2\Lambda) \rightarrow 1 $. Thus, in the large $k$ region, there is a $m_\chi^2$ dependence of the density contrast since $\tilde{\delta}_{\chi}(k,\eta)\propto k_{\rm FS}^{2}\propto m_\chi^2$. 

In summary, we derive the asymptotic expansion of \cref{eq:master_eq} in the limit of large and small momenta (or wavenumber) for bosons,
\begin{equation}
    \delta \tilde{n} _\chi(k,\eta)\simeq   \frac{{\bar n}_\chi}{\bar{\rho}_{\rm CDM}}\delta \tilde{\rho }_{\rm CDM}(k,\eta)
    \left\{\begin{array}{lr}1 & ~~k \ll k_{\rm FS} \\ \ln (2\Lambda)k_{\rm FS}^2/k^2 &~~ k \gg k_{\rm FS} \end{array}\right.\,.
\end{equation}
For fermions, the expression is the same with setting $\ln (2\Lambda)\rightarrow1$.

\bibliographystyle{JHEP}
\bibliography{ref}
\end{document}